\documentclass[12pt,a4paper]{article}
\usepackage{jheppub_kim}
\usepackage{slashed}
\usepackage{longtable}
\usepackage{epsfig}
\usepackage{amssymb,amsmath}
\usepackage{amsthm}
\usepackage{graphicx}
\usepackage{graphics}
\usepackage[hang,nooneline,scriptsize]{subfigure}
\usepackage{subfigure}
\usepackage{array}
\usepackage{color,soul}
\usepackage{wasysym}
\begin{document}
  
\title{Leading-order corrections to the thermodynamics of Rindler modified Schwarzschild black hole}
 
\author[a,b]{Surajit Mandal,}
\author[c]{Surajit Das,}
\author[d,e]{Dhruba Jyoti Gogoi,\footnote{Corresponding author}}
\author[f]{Ananda Pramanik}

\affiliation[a]{Department of Physics, Jadavpur University, Kolkata, West Bengal 700032, India.}
\affiliation[b]{Department of Physics, AKPC Mahavidyalaya, Bengai, West Bengal 712611, India.}
\affiliation[c]{Department of Physics, Cooch Behar Panchanan Barma University, Vivekananda Street, Coochbehar, West Bengal 736101, India.}
\affiliation[d]{Department of Physics, Dibrugarh University, Dibrugarh, Assam 786004, India.}
\affiliation[e]{Theoretical Physics Division, Center for Atmospheric Studies, Dibrugarh University, Dibrugarh, Assam 786004, India.}
\affiliation[f]{Department of Mathematics, Belda College, Paschim Medinipur, West Bengal 721424, India.}

\emailAdd{surajitmandalju@gmail.com}
\emailAdd{surajit.cbpbu20@gmail.com} 
\emailAdd{moloydhruba@yahoo.in}
\emailAdd{anandapramanik100@gmail.com}

\abstract{In this work, we present a thermodynamical study of a Rindler modified Schwarzschild black hole under the consideration of small thermal fluctuations. In particular, we compute various stable macroscopic thermodynamic variables such as Hawking temperature, entropy, Helmholtz free energy, internal energy, enthalpy, and Gibbs free energy. To explore the effects of small statistical thermal fluctuations on stable thermodynamical parameters, we estimated the corrections to the various thermodynamical potentials of Rindler modified Schwarzschild black hole up to the first (leading) order and do a comparative study for the different values of correction parameter and Rindler acceleration parameter for fixed values of a cosmological constant. \textcolor{black}{In this study, we examine the stability of black holes in the presence of thermal fluctuations. We find that when the correction parameter is positive, small-sized black holes remain stable, while large-sized ones become unstable. Conversely, when the correction parameter is negative, both small and large black holes exhibit instability. Additionally, we demonstrate that the first law of thermodynamics remains valid even in the presence of thermal fluctuations.} }


\maketitle

\section{Introduction}
In 1973, it was introduced by Bardeen, Carter, and Hawking that a black hole could be treated as a thermodynamic object and the dynamics of a black hole should be summarized by four laws which are basically analogous to the laws of thermodynamics \cite{m1}. In view of the classical theory of general relativity, this hypothesis flourished on a logical analogy where the area of event horizon \cite{m2} (or in other words, the square of irreducible mass \cite{m3}) never decreases. The second law of thermodynamics says that the entropy of any thermodynamical system never decreases in principle. Thus in an analogical way, we have the second law of black hole thermodynamics which claims that the entropy of a black hole is proportional to the area of the event horizon. This analogy makes the beginning of a new area of research which is black hole thermodynamics. In the zeroth and first law of black hole thermodynamics, surface gravity is constant over the event horizon of the stationary black hole and it plays the role of the temperature in black hole thermodynamics. The surface gravity ($\kappa$) and temperature ($T_{H}$) are related via the relation $T_{BH}=\frac{\hbar\kappa}{2\pi}$, where $\hbar$ is the reduced Planck constant. There subsists a one-to-one resemblance between black hole thermodynamics and classical thermodynamics \cite{m4}.

In 1973, Bekenstein and Hawking established a quantitative relation among the entropy ($S$) and the area ($A$) of the event horizon of a black hole \cite{m1,m5} and they are related by the relation $S=\frac{A}{4}$. It should be accentuated here that, in thermodynamics, the role of pressure is played by the cosmological constant. At first, people suspect that the second law of thermodynamics can be violated due to the assumption that anything that is swallowed by the black hole can never emerge. Hence, a part of the information (entropy) it contains will be lost forever. A maximum entropy was assigned to save the second law of black hole thermodynamics from being violated this law \cite{m6,m7,m8}. Moreover, the employment of maximum entropy to a black hole conveyed the discovery of holographic duality \cite{m9,m10}. This illustrates the degrees of freedom in space-time along its boundary surface and when the size of a black hole decreases, due to small thermal fluctuations around equilibrium, the area law will then be corrected. How entropy can be modified due to these thermal fluctuations is a very rigorous question. These thermal fluctuations appear from quantum corrections to the black hole geometry delineate the system under consideration.

In order to get the answer to this question, people developed different approaches but as a consequence, they arrive at the same decision that the leading-order modification to a small-sized black hole is logarithmic in nature. For instance, logarithmic corrections to the entropy of a black hole were studied by considering Rademacher expansion \cite{m11}. In 1997, Sergey N. Solodukhin also demonstrated the same results in the case of string-black hole correspondence \cite{m12}. By taking into account of thermal fluctuations, the correction to entropy gives rise to the modification in the different thermodynamical equations of states and found in Refs. \cite{m13,m14,m15,m15a,m16,m17,m18,m19,m20}. How small thermal fluctuations can affect the rotating and charged BTZ black hole, massive black hole in Ads space, and the Godel black hole were studied in Refs. \cite{m21,m22,m23,m24}. Furthermore, the first-order corrections to the dilatonic black hole \cite{m25} and Schwarzchild Beltrami-de sitter black hole \cite{m26} were analyzed. Using the partition function approach, the effect of thermal fluctuations near the equilibrium of a small black hole was also emphasized \cite{m27}. A. Chamblin et al. have studied the holography, thermodynamics, and fluctuations for the charged Reissener-Nordstr$\ddot{o}$m black hole \cite{m28}. A study by Hawking and Page resembles that AdS Schwarzschild's black hole undergoes a phase transition when it reaches a critical temperature \cite{m29}. This results in the establishment of the study of the $P-V$ criticality of black holes. Later on, Hawking’s work has been verified in other complicated Ads space-times also \cite{m30,m31}. Such investigations give an analogy between the Van der Walls liquid-gas system and the charged black hole. Further, B. Pourhassan et al. have studied some thermodynamic properties of a black hole using an adaptive model of graphene \cite{m32} and have also studied quantum fluctuations of BTZ black hole in massive gravity \cite{m33}. In Ref. \cite{m34},  S. Chougule et al. have studied Hawking radiation and black hole chemistry by using tunneling formalism. How thermal fluctuations can affect the thermodynamics of black holes considering a hyperscaling violation has been analyzed in Ref. \cite{m19}. Study on the lower bound violation of shear viscosity to entropy ratio based on the logarithmic correction in STU model found in \cite{m35}. The stability of the STU black hole is changed due to these corrections which came from thermal fluctuations. As a consequence of thermal fluctuations, the effect of first-order corrections to a dumb hole (which is an analogue to a black hole) have been studied in Ref. \cite{m36}. Furthermore, how statistical thermal fluctuations can affect the thermodynamics of a charged dilatonic black Saturn and singly spinning Kerr-Ads black hole was studied by  B. Pourhassan et al. \cite{m37,m38}. However, it was found from the study of logarithmic corrections to the thermodynamics of the modified Hayward black hole that, quantum fluctuations do not affect the stability of this black hole \cite{m39}. Lately, the elevation of quantum fluctuations from the small thermal fluctuations is introduced in \cite{m40}. \textcolor{black}{ Apart from these studies, there are several significant studies dealing with different properties of black holes in different theories of gravity \cite{Banerjee1, Banerjee2, Lambiase01,dj, Banerjee3, Banerjee4, Banerjee5}. }

In 2011, an effective model was constructed for the gravity of a focal object at a large distance by Grumiller \cite{r1}. In Grumiller’s modified gravity, the effective potential includes the Newtonian potential as well as a Rindler term. The resulting metric is known as
the Rindler modified Schwarzchild black hole (RMSBH) \cite{r1a}. Initially, RMSBH aims to explain the mysterious acceleration that acts on the Pioneer spacecraft \cite{r1b}. However, Rindler’s acceleration term results in the force law which has a considerable effect on gravity at very large distances. Moreover, Turyshev et al. \cite{r1c} have recently done an alternative study to Grumiller’s modified gravity where the Pioneer anomaly is described by thermal heat loss of the satellites. However, RMSBH can provide a theoretical explanation of a few issues such as rotation curves of spiral galaxies, perihelion shift in planetary orbits, and the gravitational redshift. The rotation of the curve for the local galaxies can be explained by using the Rindler acceleration in quantum-corrected gravity theories and can be found in Refs. \cite{r1,r1g,r1b}. As an alternative to the dark matter in galaxies, we studied the new Rindler acceleration term and eventually, it is checked by considering the HI Nearby Galaxy Survey (using 8 galaxies) which results in the Rindler acceleration parameter of around $a \approx 3\times 10^{-9} cm/s^2$ \cite{r1h,r1i,r1j,r1k}. A study of quantum tunneling and spectroscopy of the area/entropy of RMSBH can be found in Refs. \cite{r1e,r1f}. Recently, I. Sakalli et al. have studied on Hawking radiation and deflection of light for RMSBH \cite{r2}. In their study, they showed the effect of the Rindler acceleration on the deflection of light. Here, in the present study, we aim to study the modified (quantum-corrected) thermodynamics of RMSBH. In this regard, we will explore how the Rindler acceleration parameter affects various thermodynamical parameters.

\textcolor{black}{In this work, we examine a Schwarzschild black hole modified by the Rindler acceleration term and analyze its thermal properties, including the impact of thermal fluctuations. We calculate the leading-order correction to the entropy of the Rindler modified Schwarzschild black hole (RMSBH) to understand its behavior near equilibrium when stable thermal fluctuations are considered. Our analysis shows that thermal fluctuations significantly affect the entropy of small-sized black holes but have a lesser impact on larger ones. Additionally, we derive corrected expressions for enthalpy, Helmholtz free energy, and thermodynamic volume, ensuring compliance with the first law of thermodynamics. With these expressions, we calculate the corrected internal energy and Gibbs free energy.}

%This work assumes a Schwarzschild black hole which is modified by the Rindler acceleration term and discusses their various thermal properties. Moreover, the effects of thermal (statistical) fluctuations on the thermodynamics of this black hole have also been studied. To get the answer to the question of what happens in the vicinity of the equilibrium when small stable fluctuations of the thermal system are taken into account, we compute the leading-order correction to the entropy of Rindler modified Schwarzschild black hole (RMSBH). After that, we present a graphical analysis of the entropy with respect to the event horizon radius for two cases with and without taking into account of thermal fluctuations. Here, we notice that the thermal fluctuations affect the entropy for small-sized black holes significantly and their impacts are less for large black holes. Furthermore, we calculate the corrected enthalpy energy and Helmholtz free energy of the system with due help of the Hawking temperature and corrected entropy. The pressure should be expressed in terms of the cosmological constant. Hence, the corrected thermodynamic volume of RMSBH is evaluated using the expression of corrected enthalpy energy. This can be done due to the fact that the system should satisfy the first law of thermodynamics. Once we have an expression for the corrected enthalpy energy, Helmholtz free energy, and volume, it is a matter of calculation to find corrected internal energy and Gibbs free energy. 

The main goal of this work is to study correction on various thermodynamic parameters of RMSBH as soon as small statistical fluctuations around the equilibrium point are taken into consideration. We will also study the effect of the Rindler acceleration parameter on corrected entropy, specific heat and various perturbed thermodynamic potentials. Throughout this paper, we use the signature $(-,+,+,+)$ and consider the geometrical units (fundamental constants) $c=G=\hbar=k_{B}=1$. This work is presented as follows. In section \ref{sec1}, we give a brief overview of the geometry of RMSBH. In section \ref{sec2}, we study the event horizon of RMSBH. Within this section, we also study Hawking temperature, uncorrected entropy, Helmholtz free energy, enthalpy, volume, internal energy, and Gibbs free energy. The effects of thermal fluctuations on entropy have been studied in \ref{sec3} and derive various leading-order corrected thermodynamic parameters in section \ref{sec4}. In section \ref{sec5}, we study the stability and the validation of the first law of thermodynamics of RMSBH under the effect of thermal fluctuations. Finally, we summarize our results and final remarks in section \ref{sec7}.

\section{Brief review of Rindler modified Schwarzschild black hole geometry}\label{sec1}
In this section, we will discuss briefly on the geometry of Rindler-modified Schwarzschild black hole. An effective model was constructed by Grumiller \cite{r1} for the gravity of a central object at outside the galaxy (large scale), usually known as RMSBH geometry. The presence of the Rindler term in the RMSBH spacetime creates an anomalous acceleration in the geodesics of test particles. For describing the dynamics of this theory under the consideration of spherically symmetric line element of four-dimensional spacetime having the form:
\begin{equation}\label{00}
ds^2=g_{ab}(x^i)dx^{a}dx^{b}+\Phi^2(x^i)(d\theta^2+\sin^2\theta d\phi^2),
\end{equation}
the generic effective theory of dilation gravity gives us RMSBH solution and is described by using the process of spherical reduction \cite{s1} from four-dimensional to a specific two-dimensional action as the following \cite{s2,s3,r2}:
\begin{equation}\label{1}
\mathcal{A}=-\int d^2x\sqrt{-g}[\Phi^2R+2(\partial\Phi)^2+8a\Phi-6\Lambda\Phi^2+2],
\end{equation}
where $g=det(g_{ab})$ and the scalar field $\Phi$ are the objects of intrinsically two dimension ($a,b=0,1$ and $x^i={t,r}$). $\Lambda$ and $R$ represent the cosmological constant and the Ricci scalar respectively and $a$ is the Rindler acceleration \cite{s4}, which is genuinely absent in Einstein's GR. Now, taking into account the variational principle to action, Eq.(\ref{1}) and solving the corresponding field equations with a cosmological constant, one can finally appear to the following static spherically symmetric line-element for RMSBH spacetime:
\begin{equation}\label{2}
ds^2=-f(r)dt^2+\frac{dr^2}{f(r)}+r^2(d\theta^2+sin^2\theta d\phi^2),
\end{equation}
where the function $f(r)$ is defined as:
\begin{equation}\label{3}
f(r)=1-\frac{2M}{r}-\Lambda r^2+2ar.
\end{equation}
In Eq.(\ref{3}), the quantity $M$ represents an integral constant that is related to the mass of RMSBH. One can easily recover the Schwarzschild solution if $a=\Lambda=0$. One may note that with $M = \Lambda = 0$, the above line-element, Eq.(\ref{2}) reduces to the two-dimensional Rindler metric \cite{r3}. The value of $\Lambda$ is approximated to $10^{-123}$ as given by \cite{r4,r5}. Therefore, one can set it to zero ($\Lambda = 0$) for simplicity but in our present study, we will not make it so to have a more general solution. The significance of the Rindler acceleration lies in its capacity to characterize the shape of the galactic rotation curve, implying a constant acceleration directed toward the source when $a$ is taken as a positive parameter. The author in \cite{r1} also claims that an apparently radial constant acceleration, which is associated with the trajectories of the Pioneer spacecraft, of order $a\approx 10^{-11}~ms^{-2}$ (known as ‘‘Pioneer anomaly’’) may be explained by the term ‘‘Rindler acceleration’’ \cite{s5,s6}. The fact behind this is that, for large distances, r of the order of the Hubble length, the Rindler term $2ar$ and the cosmological constant term $\Lambda r^2$ become more relevant and approach unity for $a\approx 10^{-10}-10^{-11}~ms^{-2}$. The same value is also for the Pioneer acceleration and the MOND characteristic acceleration \cite{s7}. It is worth mentioning here that it does not necessarily spoil the solar-system precision tests as the Rindler acceleration $a$ is system independent in this effective dilation scalar field model of gravity.

However, at the IR region\footnote{we use at a large distance limit as ‘‘infrared’’, abbreviated by IR by the convention of the particle physicists’ jargon.} the value of the Rindler acceleration is estimated as $a \approx 10^{-62}-10^{-61}~ms^{-2}$ \cite{r1}. Without loss of generality, due to the small value of $a$, to observe the effects of $a$ on various thermodynamical parameters by taking its different values, all the graphs are scaled with $a$ taking values of order 1, 4, 6 and this is just a scale choice for presenting the graphical behavior of the results.
 
Very recently, it was shown by Mazharimousavi and Halilsoy (MH) \cite{r8} that the RMSBH metric becomes physically acceptable in $f(R)$ gravity framework. The metric of \cite{r8} is familiar as Grumiller–Mazharimousavi–Halilsoy black hole. Recently, a study on the geodesics of this black hole was found in \cite{r1d}. Here, we should mention that the RMSBH solution describes an uncharged black hole for a central object at a large scale which corroborates with the Grumiller–Mazharimousavi–Halilsoy black hole (GMHBH) in the absence of cosmological constant $\Lambda$. The present analysis of modified thermodynamics for the uncharged RMSBH will be equally applicable to the case of Mazharimousavi–Halilsoy black hole because the solution, Eq.(\ref{3}) does coincide with the Mazharimousavi–Halilsoy black hole in the cosmological constant less limit.

Rindler acceleration has recently become furore again for an observer accelerated in a flat spacetime \cite{r9}. At first, Grumiller \cite{r1} and then jointly with his collaborators \cite{r1g,r10} have exposed the correlation between the Rindler acceleration and the Pioneer anomaly. Recently, the effect of the Rindler acceleration parameter $a$ on the Hawking temperature was studied by S.F. Mirekhtiary et al \cite{r1e}. In 2013, S.H. Mazharimousavi et al. showed that $a$ plays the role of dark matter \cite{r8}. Here, in the present work along with the corrected thermodynamics, we also aim to study the role of the Rindler acceleration parameter $a$ on various thermodynamical parameters.

\begin{figure}[ht]
\begin{center}
\includegraphics[width=0.8\linewidth]{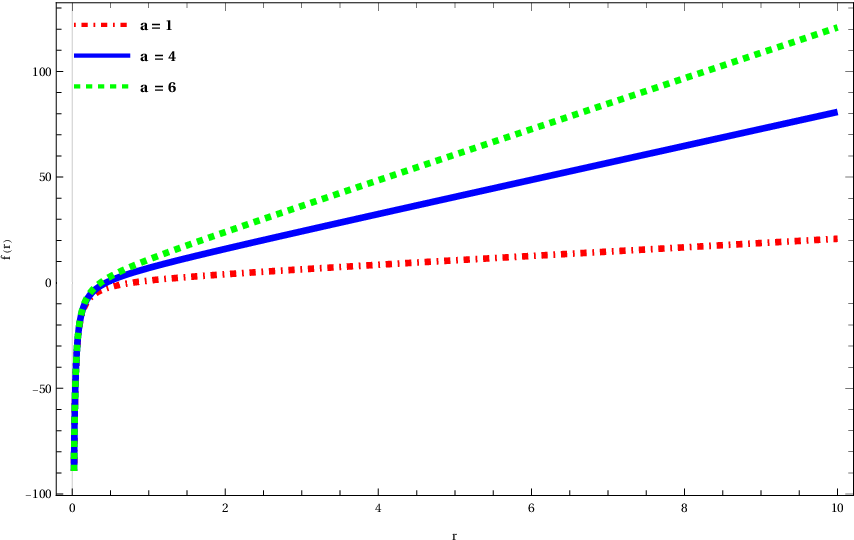}
\end{center}
\caption{Plot of lapse function $f(r)$ versus $r$ for differents $a$ with $M=1$, $\Lambda=10^{-123}$.}
\label{fig1}
\end{figure}
 
Fig.\ref{fig1} shows the changes in $f(r)$ with $r$ for three different values of $a$. By increasing the Rindler acceleration ($a$) of RMSBH, the lapse function $f(r)$ significantly increases for all states.

\section{Event horizon and thermodynamical study}\label{sec2}
In this section, we will discuss about event horizon as well as the thermodynamics of RMSBH. The event horizon radius $r_{+}$ of a black hole can be calculated by taking $f(r) = 0$. Interestingly the computation reveals that among the three different roots of $r$, only one becomes the real positive root which in turn provides us the physical radius of the horizon of RMSBH. The real positive root becomes
\begin{equation}\label{4}
r_{+}=\frac{2a}{3\Lambda}+\frac{2^{\frac{1}{3}}(-4a^2-3\Lambda)}{3\Lambda A}-\frac{A}{3\times 2^{\frac{1}{3}}\Lambda},
\end{equation}
where $A=\sqrt[3]{-16a^3-18a\Lambda+54M\Lambda^2+\sqrt{4(-4a^2-3\Lambda)^3+(-16a^3-18a\Lambda+54M\Lambda^2)^2}}$. It is noticed that in the limit $\Lambda\rightarrow 0$ and $a\rightarrow 0$, we get $r_{+}=2M$, as expected.

Using the given lapse function $f(r)$ i.e., Eq.(\ref{3}), Hawking temperature $T_{H}$ can be derived as follows,
\begin{equation}\label{5}
T_{H}=\frac{f^\prime(r)}{4\pi}\Bigg|_{r\ = \ r_{+}}=\frac{1}{2\pi}\Big(\frac{M}{r^2_{+}}-\Lambda r_{+}+a\Big),
\end{equation}
where prime “$\prime$” represents the differentiation with respect to $r$ and $r_{+}$ denotes the horizon radius of a black hole. It is seen from the above expression that when the RMSBH loses its mass ($M\rightarrow 0$) as a consequence of the Hawking radiation, $T_{H}$ increases (i.e., $T_{H}\rightarrow\infty$) in a way such that the divergence speed of this black hole is modulated by $a$. However, we can observe that 
\begin{equation}\label{5a}
\lim_{(\Lambda, a)\to(0,0)} T_{H} = \frac{1}{8\pi M}
\end{equation}
which is familiar Hawking temperature calculated for the Schwarzschild black hole (SBH).

\begin{figure}[ht]
\begin{center} 
\includegraphics[width=0.7\linewidth]{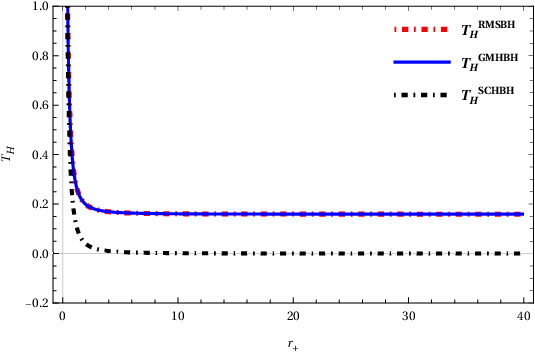}
\end{center}
\caption{Comparison of variation of the Hawking temperature against the horizon radius $r_{+}$ with $a=1$, $\Lambda=10^{-123}$ for RMSBH, $a=1$, $\Lambda=0$ for GMHBH and $a=0$, $\Lambda=0$ for SBH, respectively. (Here, for all plots, we set $M=1$.)}
\label{fig1a}
\end{figure}

The characteristics of Hawking temperature with horizon radius are presented in Fig.\ref{fig1a}. It is obvious from this plot that the temperature is a decreasing function of $r_{+}$ and curved is positively valued. Fig.\ref{fig1a} shows the comparison of Hawking temperature against horizon radius for the RMSBH, GMHBH, and SBH. The plot depicts that the temperature behavior for all the above black holes take the same nature i.e. the temperature is always positive. Interestingly, the temperature curve for the RMSBH case coincides with the GMHBH case due to the small value of a cosmological constant, as expected.

\begin{figure}[ht]
\begin{center}
\includegraphics[width=0.8\linewidth]{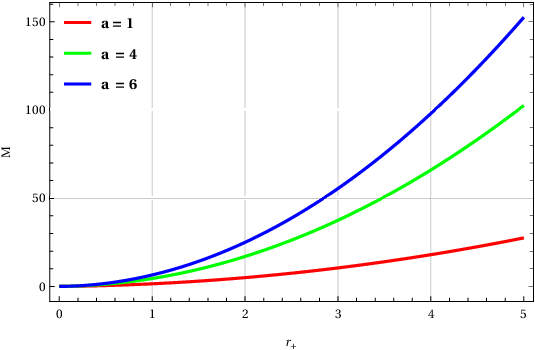}
\end{center}
\caption{Variation of physical mass of black hole with horizon radius $r_{+}$ for $a=1,~4~\&~6$ with $\Lambda=10^{-123}$.}\label{fig1b}
\end{figure}

\textcolor{black}{The behavior of physical mass with horizon radius for different values of $a$ is depicted in Fig.\ref{fig1b}. The figure indicates that the mass of RMSBH is an increasing function of horizon radius and mass increases when $a$ increases.}

Now, the corresponding Bekenstein entropy, using the four fundamental laws of black hole thermodynamics, is calculated by
\begin{equation}\label{6}
S_{0}=\pi r^2_{+}.
\end{equation}
Now, using given Hawking temperature Eq.(\ref{5}) and entropy Eq.(\ref{6}), the expression for Helmholtz free energy (represented by $F_{r}$ and subscript r is used here to indicate the free energy of RMSBH) of RMSBH can be estimated as follows:
\begin{equation}\label{7}
F_{r}=-\int S_{0}~dT_{H}=\frac{r_{+}}{4}(1+\Lambda r_{+}^2).
\end{equation}
Moreover, we can compute the enthalpy energy ($H$) of the system using the following standard formula:
\begin{equation}\label{8}
H=\int T_{H} dS_{0}.
\end{equation}
Substituting the corresponding values of temperature (\ref{5}) and entropy (\ref{6}) in the above equation we have,
\begin{equation}\label{9}
H=\frac{r_{+}}{2}(1+2ar_{+}-\Lambda r_{+}^2).
\end{equation}
This expression is for the enthalpy energy of RMSBH. The expression for pressure is calculated as
\begin{equation}\label{10}
P=-\frac{\Lambda}{8\pi}.
\end{equation}
Writing $H$ in terms of $P$, we attains
\begin{equation}\label{11}
H=\frac{r_{+}}{2}(1+2ar_{+}+8\pi P r_{+}^2).
\end{equation}
Now, the thermodynamic volume is obtained as
\begin{equation}\label{12}
V=\frac{dH}{dP}=4\pi r^3_{+}.
\end{equation}
In order to calculate the further properties of the system such as internal energy ($U$) and Gibbs free energy ($G_{r}$), we need the above-derived quantities. Nevertheless, it is familiar that internal energy ($U$) is derived mathematically by the following relation
\begin{equation}\label{13}
U=H-PV.
\end{equation}
Plugging the corresponding values of enthalpy energy, pressure and volume in Eq.(\ref{13}), we get
\begin{equation}\label{14}
U=\frac{r_{+}}{2}(1+2ar_{+}).
\end{equation}
This expression refers to internal energy of RMSBH. In the same manner, we can derive the expression for Gibbs free energy ($G_{r}$). The thermodynamical formula for Gibbs free energy is given by
\begin{equation}\label{15}
G_{r}=F_{r}+PV.
\end{equation}
For a given values of $F_{r}$, $P$ and $V$, the Gibbs free energy is calculated by
\begin{equation}\label{16}
G_{r}=\frac{r_{+}}{4}(1-\Lambda r_{+}^2).
\end{equation}
In the following sections, we would like to observe the effect of small stable fluctuations near equilibrium on the thermodynamics of RMSBH.

\section{The leading-order corrections to entropy}\label{sec3}
The entropy of a thermal system can be calculated in two ways: 

\LEFTcircle \ By taking into consideration a microcanonical ensemble where an equilibrium system is considered as truly isolated; 

\LEFTcircle \ By taking into account a canonical ensemble such that at a fixed temperature, the system is medicating as being in thermal contact with a very large reservoir. 

The canonical ensemble discretion cannot be a good alternative for the systems like strongly self-gravitating black holes at equilibrium. Compared to the former case, the latter will be useful for our present study as the consideration of canonical ensemble works very well for the small-sized statistical (thermal) system. We calculate the entropy of RMSBH due to small statistical fluctuations around equilibrium. In order to calculate entropy correction, we ﬁrst deﬁne general partition function for the thermal system of canonical black hole as
\begin{equation}\label{17}
Z(\beta)=\int_{0}^{\infty}\rho(E)e^{-\beta E}dE,
\end{equation}
where $\beta=\frac{1}{T_{H}}$. For convenience, we set the Boltzmann constant, $k_{B}$ to unit here. Now, for a given partition function, the density of state $\rho(E)$ can be calculated using the inverse Laplace transformation of Eq.(\ref{17}) as \cite{r11,r12}
\begin{equation}\label{18}
\rho(E)=\frac{1}{2\pi i}\int_{\beta_{0}-i\infty}^{\beta_{0}+i\infty}e^{\ln Z(\beta)+\beta E}d\beta.
\end{equation}
Here one should mention that the exponential term represents the exact entropy of a black hole. So it can be written as
\begin{equation}\label{19}
S(\beta)=\ln Z(\beta)+\beta E.
\end{equation}
This explicit temperature-dependent exact entropy is the total sum of entropies of all individual sub-systems. Thermal fluctuations play a very crucial role only when the size of the black hole is very small (in other words, when the event horizon radius is very small) and this also resembles the canonical ensemble consideration. In order to observe the effects of thermal (small) fluctuations on entropy one can expand $S(\beta)$ around equilibrium $\beta=\beta_{0}$ (via Taylor expansion) as follows
\begin{equation}\label{20}
S(\beta)=S_{0}+\frac{1}{2}(\beta-\beta_{0})^2\frac{\partial^2S}{\partial\beta^2}\Bigg|_{\beta=\beta_{0}}+\text{higher order terms},
\end{equation}
where $S_{0}(=S(\beta_{0}))$ denotes the canonical entropy for saddle point equilibrium (temperature). Putting the above value of entropy from Eq.(\ref{20}) in Eq.(\ref{18}), we obtain
\begin{equation}\label{21}
\rho(E)=\frac{e^{S_{0}}}{2\pi i}\int_{\beta_{0}-i\infty}^{\beta_{0}+i\infty}exp\Bigg[{\frac{1}{2}(\beta-\beta_{0})^2\frac{\partial^2S}{\partial\beta^2}}\Bigg|_{\beta=\beta_{0}}\Bigg]d\beta.
\end{equation}
On simplification of the above integral equation, we arrive at the following density of states as
\begin{equation}\label{22}
\rho(E)=\frac{e^{S_{0}}}{\sqrt{2\pi\frac{\partial^2S}{\partial\beta^2}\Big|_{\beta=\beta_{0}}}}.
\end{equation}
The logarithm of the above density of states leads to the corrected entropy due to thermal fluctuations and is obtained as
\begin{equation}\label{23}
S(\beta)=\ln \rho(E)=S_{0}-\frac{1}{2}\ln\frac{\partial^2S}{\partial\beta^2}\Bigg|_{\beta=\beta_{0}}+\text{sub-leading terms}.
\end{equation}
Here, we have neglected the higher-order correction terms. This relation can play a significant role for all thermodynamic systems such as the black hole system and can be considered as a canonical ensemble. The correction term is calculated by \cite{r6}
\begin{equation}\label{24}
\frac{\partial^2S}{\partial\beta^2}\Bigg|_{\beta=\beta_{0}}=BT_{H}^2,
\end{equation}
where $B$ refers to the dimensionless specific heat. This helps us to obtain the micro-canonical entropy at equilibrium as
\begin{equation}\label{25}
S(\beta)=\ln \rho(E)=S_{0}-\frac{1}{2}\ln (BT_{H}^2)+\text{sub-leading terms}.
\end{equation}
For no working system at saddle point $S_{0}$, in the case of RMSBH, the specific heat coincides with the equilibrium value of canonical entropy. This leads to the leading-order entropy at equilibrium as
\begin{equation}\label{26}
S(\beta)=S_{0}-\frac{1}{2}\ln (S_{0}T_{H}^2).
\end{equation}
Without loss of any generality, we replace the factor $\frac{1}{2}$ by a more general correction parameter “$\alpha$” in the second term of Eq.(\ref{25}). The most general form of corrected entropy becomes
\begin{equation}\label{27}
S(\beta)=S_{0}-\alpha \ln( S_{0}T_{H}^2).
\end{equation}

\begin{figure}[ht]
\begin{center} 
$\begin{array}{cccc}
\subfigure[]{\includegraphics[width=0.52\linewidth]{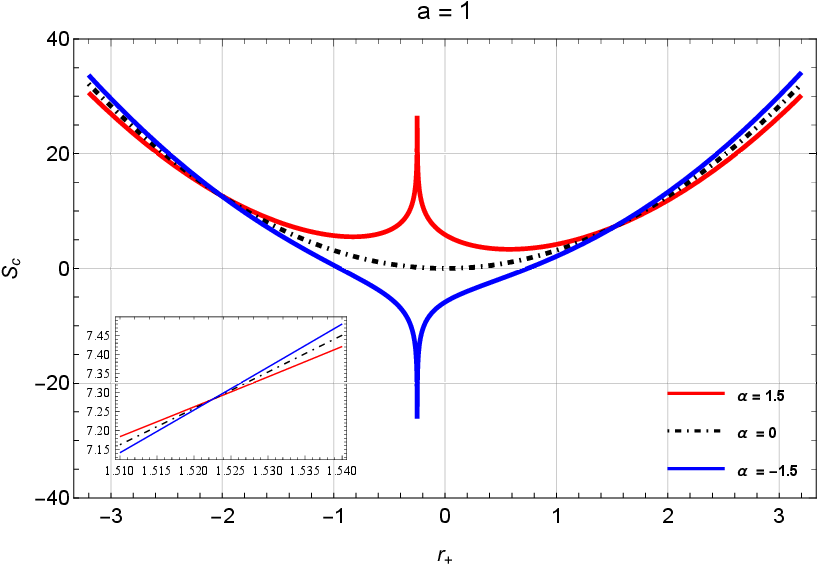}\label{2a}}
\subfigure[]{\includegraphics[width=0.52\linewidth]{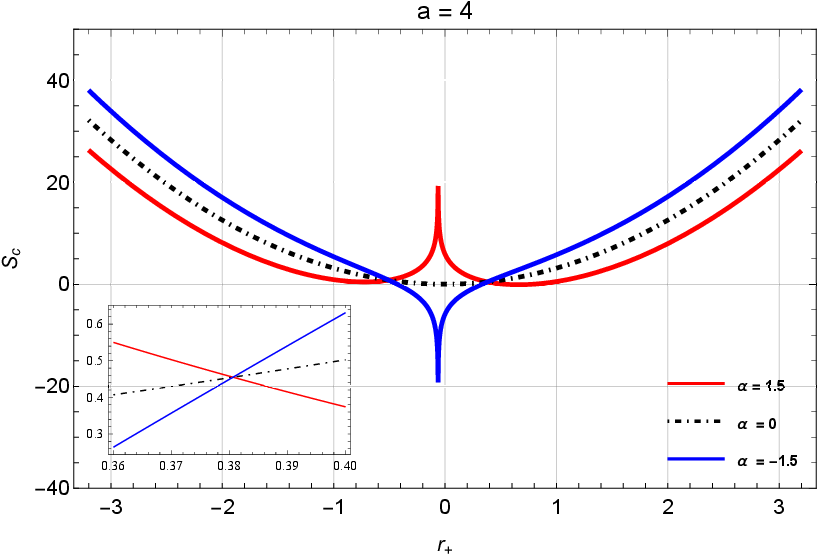}\label{2b}}\\
\subfigure[]{\includegraphics[width=0.52\linewidth]{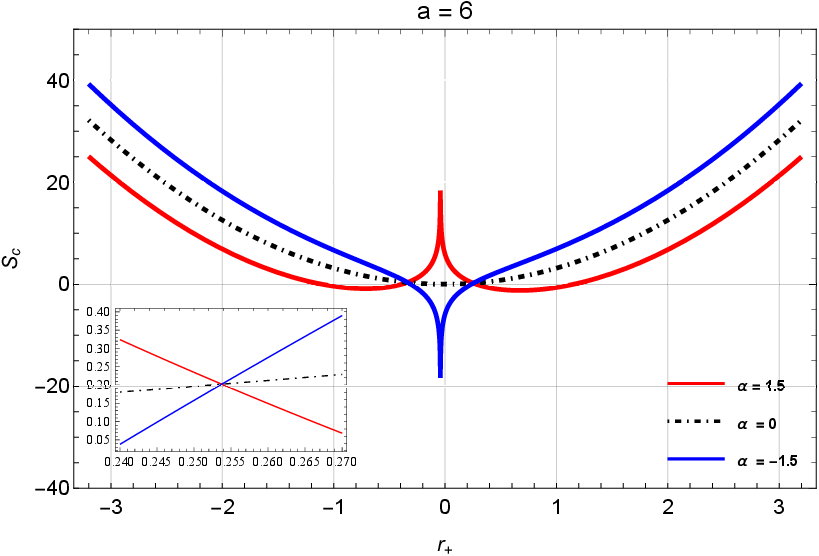}\label{2c}}
\end{array}$
\end{center}
\caption{Variation of entropy with the black hole horizon radius $r_{+}$. Here, $\alpha=0$ (no correction) is denoted by a black dash-dot line, $\alpha=1.5$ (with correction) is denoted by a red curve, and $\alpha=-1.5$ (with correction) is denoted by a blue curve. We set $\Lambda=10^{-123}$.}
\label{fig2}
\end{figure}

\textcolor{black}{The effect of thermal fluctuations on the entropy of the system is characterized by $\alpha$ which was introduced by hand in Ref. \cite{rev01} to track the corrected terms.} It is notified that the corrected entropy has a logarithmic dependency. At the saddle point, for our considered black hole system (RMSBH), the canonical entropy is determined from Bekenstein-Hawking area law \cite{r7}. Plugging the values of Hawking temperature (from Eq.(\ref{5})) and Bekenstein-Hawking area law entropy (from Eq.(\ref{6})) in above Eq.(\ref{27}), finally we appear at the micro-canonical entropy of RMSBH at equilibrium as follows,
\begin{equation}\label{28}
S_{c}=\pi r_{+}^2-\alpha \ln\Big[\frac{(M-\Lambda r_{+}^3+ar_{+}^2)^2}{4\pi r_{+}^2}\Big]=\pi  r_{+}^2-\alpha  \ln \left[\left(4 a r_{+}-3 \Lambda  r_{+}^2+1\right)^2\right]+\alpha  \ln (16 \pi ).
\end{equation}
Here, we can notice from the above expression that the last two terms represent the correction to the entropy due to the thermal fluctuations around the equilibrium. For the limit $\alpha = 0$, the corrected entropy calculated above reduces to the equilibrium entropy of our considered black hole. A comparative plot of the corrected and uncorrected entropy at equilibrium against the event horizon radius for different values of correction parameter $\alpha$ is shown in fig.\ref{fig2}.

From the plot (Fig.\ref{fig2}), we see that in the limit $\alpha\rightarrow 0$, the uncorrected (original) entropy curve at saddle point (black curve) is an increasing function of $r_{+}$ and takes positive values only. It is worth mentioning that, the entropy for black holes which have negative mass i.e. negative event horizon radius is physically meaningless.  Moreover, for small-sized black holes, entropy shows wonderful results as long as quantum corrections are taken into account. However, we observe that the nature of entropy for large-sized black holes does not undergo any significant deformation as expected. This indicates that quantum correction to entropy has importance only when a black hole has a very small horizon radius. Also, there arises \textcolor{black}{ two critical points} in this figure. It is notified that, among the \textcolor{black}{ two critical points}, the first critical point appear for those black holes which have a negative mass which is again physically meaningless. In between the \textcolor{black}{ first and second} critical points, entropy shows a positive peak for the \textcolor{black}{ positive} value of $\alpha$, while for \textcolor{black}{ negative} value of $\alpha$, the corrected entropy shows a negative peak. The negative values of entropy do not make any physical sense and are therefore forbidden. Now for a small black hole whose horizon radius is smaller than the \textcolor{black}{ second critical} horizon radius, the micro-canonical entropy takes a positive value corresponding to a positive value of the correction parameter and therefore it makes the system more stable whereas entropy takes negative values corresponding to a \textcolor{black}{ negative value} of the correction parameter. When a black hole has a radius greater than the \textcolor{black}{ second critical} horizon radius, then micro-canonical entropy has no significant difference and it becomes an increasing function of $r_{+}$ irrespective of $\alpha$. So black hole with a larger event horizon radius justifies the insignificance of thermal fluctuation. Interestingly, for the larger Rindler acceleration parameter ($a$), the above figure depicts that \textcolor{black}{ the critical point shifted towards the small horizon radius}.

\section{The leading-order perturbed thermodynamic potentials}\label{sec4}
In this section, we aim to evaluate various thermodynamic variables for the RMSBH system due to the small thermal fluctuations at the equilibrium. In this regard, firstly we would like to derive the enthalpy of the system. Mathematically, It is defined as follows
\begin{equation}\label{29}
H_{c}=\int T_{H}dS_{c},
\end{equation}
where $H_{c}$ is the corrected enthalpy energy. Putting the above values for Hawking temperature (from Eq.(\ref{5})) and modified Bekenstein entropy (from Eq.(\ref{28})) in Eq.(\ref{29}), we get
\begin{equation}\label{30}
H_{c}= \frac{r_+}{2}-\frac{2 a \alpha  \ln \left(r_+\right)}{\pi }+a r_+^2+\frac{3 \alpha  \Lambda  r_+}{\pi }-\frac{1}{2} \Lambda  r_+^3
\end{equation}
%M\ln|r_{+}|-\frac{\Lambda r^3_{+}}{3}+\frac{ar^2_{+}}{2}+\frac{\alpha}{\pi}\Bigg(3\Lambda r_{+}-a\ln|r_{+}|-\frac{M}{2r_{+}^2}-r_{+}\Bigg).

\begin{figure}[ht]
\begin{center} 
$$\begin{array}{cccc}
\subfigure[]{\includegraphics[width=0.52\linewidth]{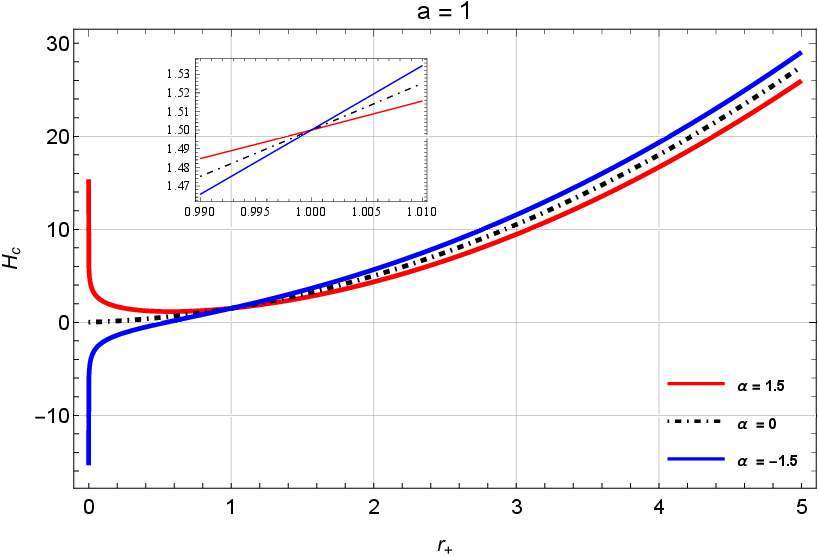}\label{3a}}
\subfigure[]{\includegraphics[width=0.52\linewidth]{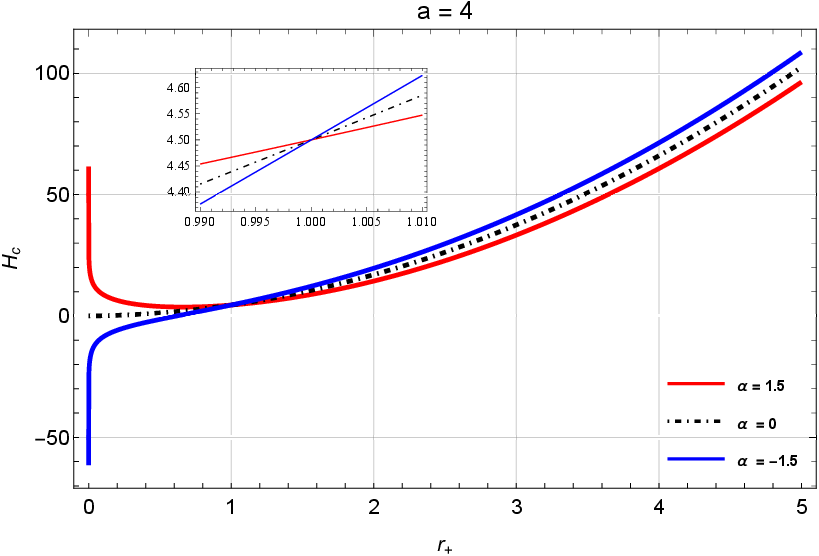}\label{3b}}\\
\subfigure[]{\includegraphics[width=0.52\linewidth]{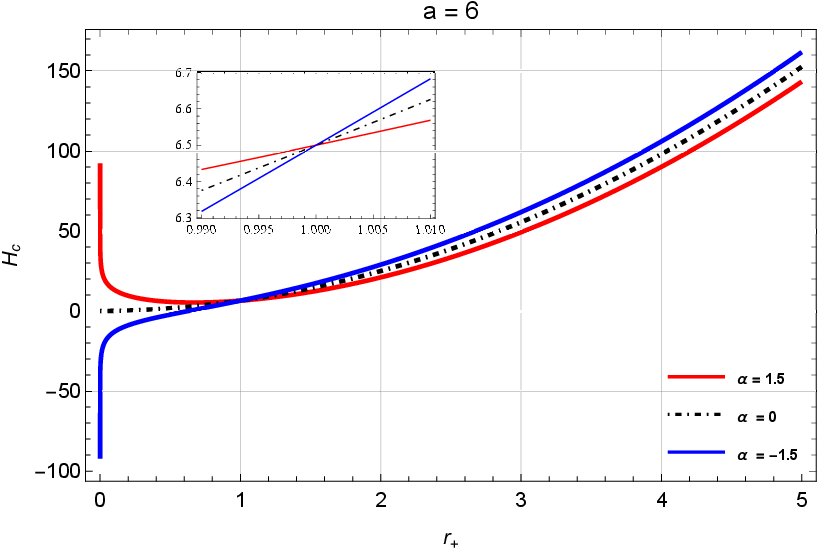}\label{3c}} 
\end{array}
$$
\end{center}
\caption{Variation of enthalpy energy with the black hole horizon radius $r_{+}$ with $\Lambda=10^{-123}$. Here, $\alpha = 0$ (no correction) is denoted by a black dash-dot line, $\alpha = 1.5$ (with correction) is denoted by a red curve, and $\alpha = -1.5$ (with correction) is denoted by a blue curve.}
\label{fig3}
\end{figure}

The above expression is for leading-order corrected enthalpy by taking into account of thermal fluctuations. This leading-order (first-order) correction term resembles the effect of thermal fluctuations near the equilibrium. Fig.\ref{fig3} depicts the behavior of resulting enthalpy energy against the event horizon radius for different values of correction parameter $\alpha$ with a fixed Rindler acceleration parameter $a$.

The plot (Fig.\ref{fig3}) shows again that thermal fluctuations will be effective only when a black hole has a small event horizon radius whereas the enthalpy of large-sized black holes will be unaffected. For black holes having a small event horizon radius, the enthalpy takes a negative value for the \textcolor{black}{ negative} correction parameter while it takes a positive value for the \textcolor{black}{ positive} correction parameter. From the plot, it is visible to us that, there exists \textcolor{black}{ a single critical point for all the cases with $a=1,4,6$. The critical point is independent of the parameter $a$. } Enthalpy energy is an increasing function of the event horizon radius for a black hole that has a larger event horizon radius.

Apart from enthalpy energy, there are three more important thermodynamic potentials to interpret the state of the system such as Helmholtz free energy, internal energy, and Gibbs free energy. We will discuss how thermal fluctuations affect these thermodynamic potentials one by one. Being a state function, free energy represents the possible amount of energy present for doing work. Helmholtz free energy has the following form:
\begin{equation}\label{31}
F_{c}=-\int S_{c}dT_{H}.
\end{equation}
Here $F_{c}$ represents the corrected Helmholtz free energy. In order to calculate free energy we plug the values of Hawking temperature and leading-order corrected Bekenstein entropy in the above expression and this leads
\begin{multline}\label{32}
F_{c}= \frac{1}{4 \pi  r_+} \Bigg[8 a \alpha  r_+ \left\lbrace \ln \left(4 a r_+-3 \Lambda  r_+^2+1\right)-\ln \left(r_+\right)\right\rbrace+\alpha  \ln (16 \pi ) \left(3 \Lambda  r_+^2-1\right)\\+\alpha  \left(1-3 \Lambda  r_+^2\right) \ln \left\lbrace \left(4 a r_+-3 \Lambda  r_+^2+1\right){}^2\right \rbrace+r_+^2 \left(12 \alpha  \Lambda +\pi  \Lambda  r_+^2+\pi \right) \Bigg]
\end{multline}
%M\ln|r_{+}|+\frac{\Lambda r^3_{+}}{6}+\frac{\alpha}{2\pi}\Bigg[-2\Big(\Lambda r_{+}-\frac{M}{r_{+}^2}\Big)\ln(M-\Lambda r_{+}^3+ar_{+}^2)+\Big(\Lambda r_{+}-\frac{M}{r_{+}^2}\Big)\ln4\pi r_{+}^2\\-8\Lambda r_{+}+4a\ln|r_{+}|-2 a\ln(M-\Lambda r_{+}^3+ar_{+}^2)-\frac{M}{r_{+}^2}\Bigg].

\begin{figure}[ht]
\begin{center} 
$$\begin{array}{cccc}
\subfigure[]{\includegraphics[width=0.52\linewidth]{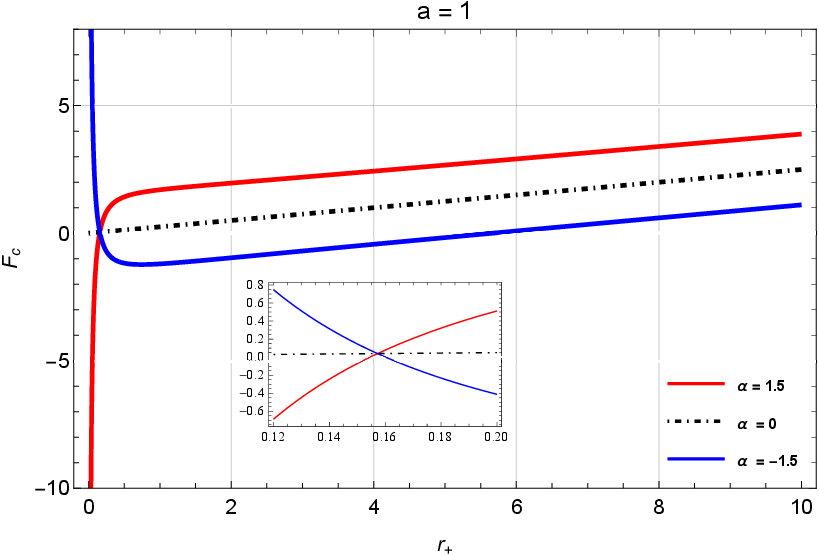}\label{4a}}
\subfigure[]{\includegraphics[width=0.52\linewidth]{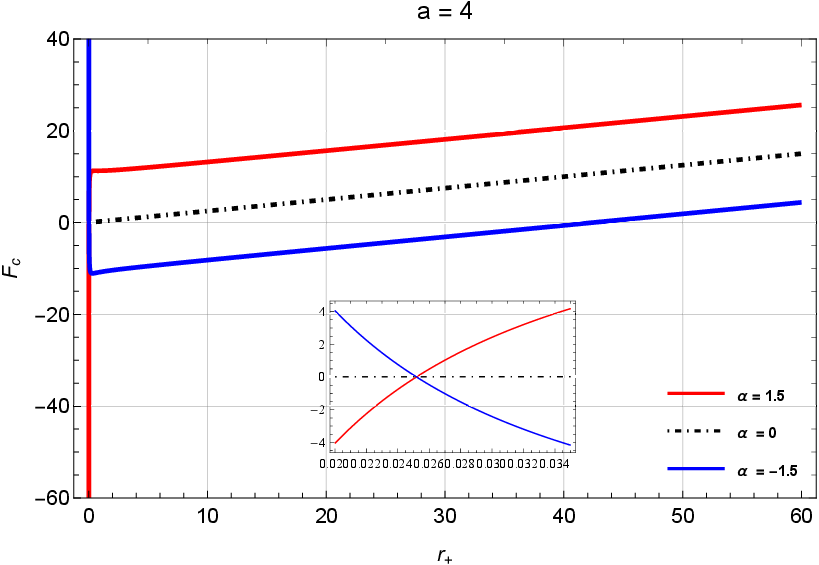}\label{4b}}\\
\subfigure[]{\includegraphics[width=0.52\linewidth]{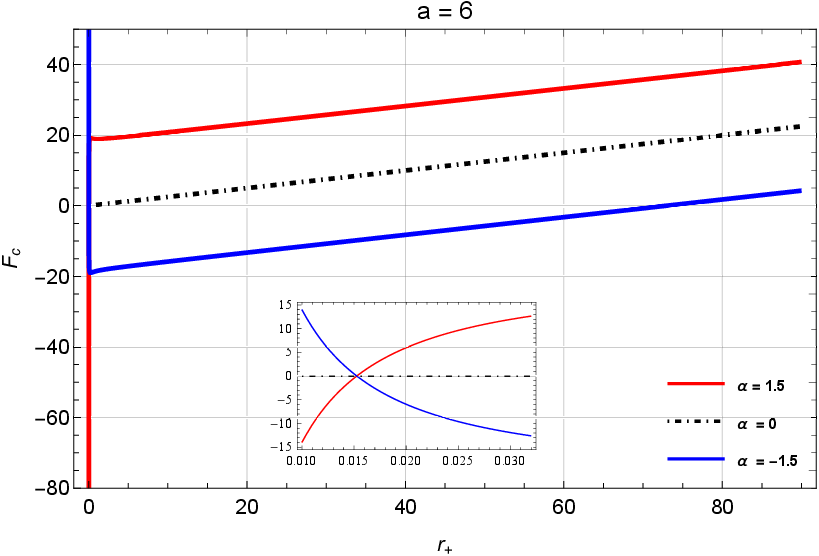}\label{4c}}
\end{array}$$
\end{center}
\caption{Variation of Helmholtz free energy with the black hole horizon radius $r_{+}$. Here $\alpha = 0$ (no correction) is denoted by a black dash-dot line, $\alpha = 1.5$ (with correction) is denoted by a red curve, and $\alpha = -1.5$ (with correction) is denoted by a blue curve. We set $\Lambda=10^{-123}$.}
\label{fig4}
\end{figure}

To study the dependency of free energy on thermal fluctuations, we plot the Helmholtz free energy vs. the event horizon radius in fig.\ref{fig4}.

The figure depicts only one critical point for free energy and this critical point occurs in the \textcolor{black}{ small black hole horizon} region. For a black hole whose horizon radius is smaller than the critical point, free energy with a positive correction parameter takes the \textcolor{black}{ negative} asymptotic value while free energy with a negative correction parameter takes the \textcolor{black}{ positive} asymptotic value. After the critical point, we notice that Helmholtz free energy shows \textcolor{black}{an increasing behavior in the positive region for the positive value of $\alpha$, while for the negative value of $\alpha$, the corrected Helmholtz free energy demeanors an increasing behavior in the negative domain initially and finally goes to the positive region by retaining the increasing nature}. All the curves \textcolor{black}{ become parallel to the uncorrected curve for large values of horizon radius. This again justifies that quantum corrections affect the thermodynamics of RMSBH significantly at small horizon radii.} The role of the larger Rindler acceleration parameter is to shift the critical point slowly towards lower $r_{+}$.

Once we know the expression for corrected enthalpy energy, we can now easily calculate the leading-order corrected volume as follows:
\begin{equation}\label{33}
V_{c}=\frac{dH_{c}}{dP}=\frac{8}{3}\pi r_{+}^3-8\alpha r_{+}.
\end{equation}

\begin{figure}[ht]
\begin{center}
\includegraphics[width=0.7\linewidth]{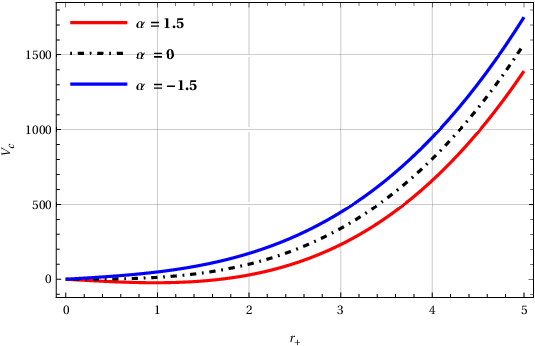}
\end{center}
\caption{Corrected volume vs. horizon radius $r_{+}$. Here, $\alpha = 0$ (no correction) represents black dash-dot line, $\alpha = 1.5$ (with correction) denotes red curve, and $\alpha = -1.5$ (with correction) denotes blue curve.}
\label{fig5}
\end{figure}

The above expression for $V_{c}$ is independent of $a$. The behavior of corrected thermodynamic volume with horizon radius is depicted in fig.\ref{fig5}. The figure shows that the correction parameter does not affect the corrected volume for small-sized as well as large-sized black holes. The figure suggests that the corrected volume of the black hole is an increasing function of the horizon radius.

\begin{figure}[ht]
\begin{center} 
$\begin{array}{cccc}
\subfigure[]{\includegraphics[width=0.52\linewidth]{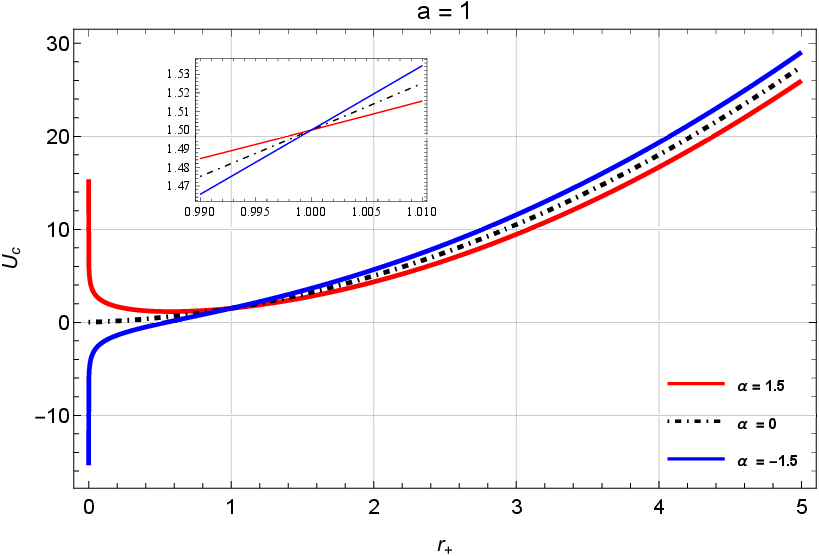}\label{6a}}
\subfigure[]{\includegraphics[width=0.52\linewidth]{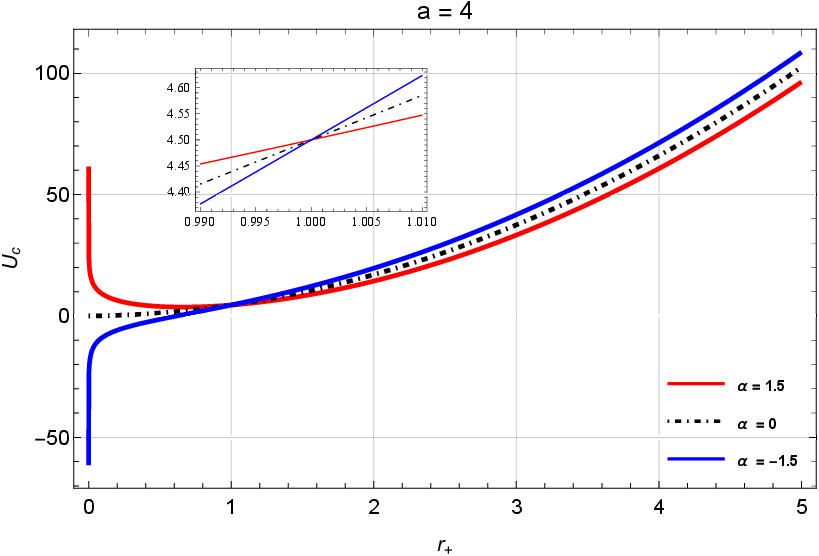}\label{6b}}\\ 
\subfigure[]{\includegraphics[width=0.52\linewidth]{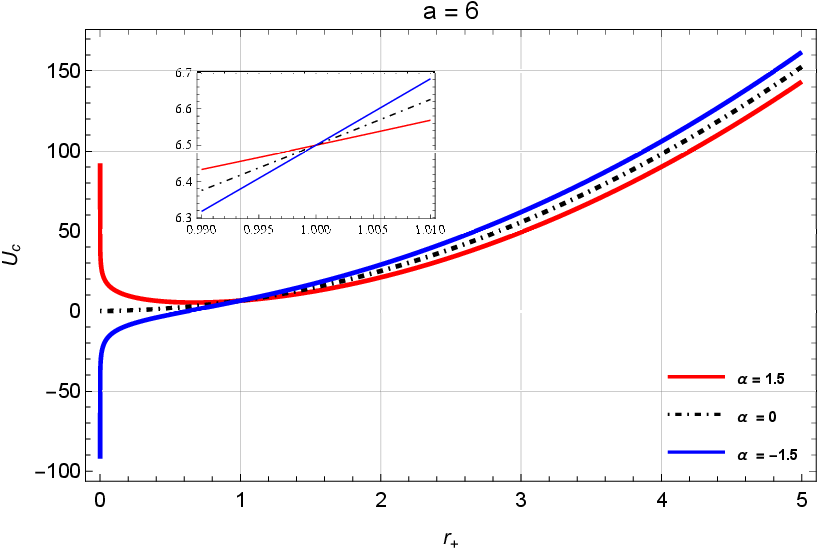}\label{6c}}
\end{array}$
\end{center}
\caption{Corrected internal energy vs. black hole horizon radius $r_{+}$ with $\Lambda=10^{-123}$. Here, $\alpha = 0$ (no correction) represents by a black dash-dot line, $\alpha = 1.5$ (with correction) is denoted by the red curve, and $\alpha = -1.5$ (with correction) is denoted by the blue curve.}
\label{fig6}
\end{figure}

Now, we would like to derive the corrected expression for internal energy. Internal energy is a state function that has inherent significance in a thermodynamical system and represents the sum of kinetic energy and potential energy ( kinetic energy appears from the motion of particles whereas potential energy arises from the particular configuration of these particles). The corrected internal energy ($U_{c}$) can be obtained from the following formula:
\begin{equation}\label{34}
U_{c}=H_{c}-PV_{c}.
\end{equation}
With the help of corrected enthalpy energy from Eq.(\ref{30}) and corrected volume from Eq.(\ref{33}), the leading order corrections to internal energy are given by,
\begin{equation}\label{35}
U_{c}= a r_+^2+\frac{r_+}{2}-\frac{2 a \alpha  \log \left(r_+\right)}{\pi }.
\end{equation}
%M\ln|r_{+}|+\frac{a}{2}r^2_{+}+\frac{\alpha}{\pi}\Big(-a \ln|r_{+}|-\frac{M}{2r_{+}^2}-r_{+}\Big)
This expression represents the effect of quantum fluctuations on internal energy. We now present the uncorrected and corrected internal energy (Eq.(\ref{35})) against the horizon radius in fig.\ref{fig6}. From the plot, it is noticed that the uncorrected internal energy goes on increasing as the size of the black hole increases. \textcolor{black}{ The internal energy changes drastically for black holes with smaller horizons than the critical point. In this regime, a positive correction parameter $\alpha$ results in positive values of internal energy while for the negative values of $\alpha$, we observe that the internal energy becomes negative. For horizon radius greater than the critical point, both cases result in positive increasing behaviour of internal energy. In this regime, the positive correction parameter values result in lower internal energy than the uncorrected one and for negative correction parameter, the internal energy is higher than the uncorrected one. An increase in the values of Rindler acceleration parameter $a$ does not shift the critical point, but it increases the internal energy significantly at higher values of $r_+$.}

\begin{figure}[ht]
\begin{center} 
$\begin{array}{cccc}
\subfigure[]{\includegraphics[width=0.52\linewidth]{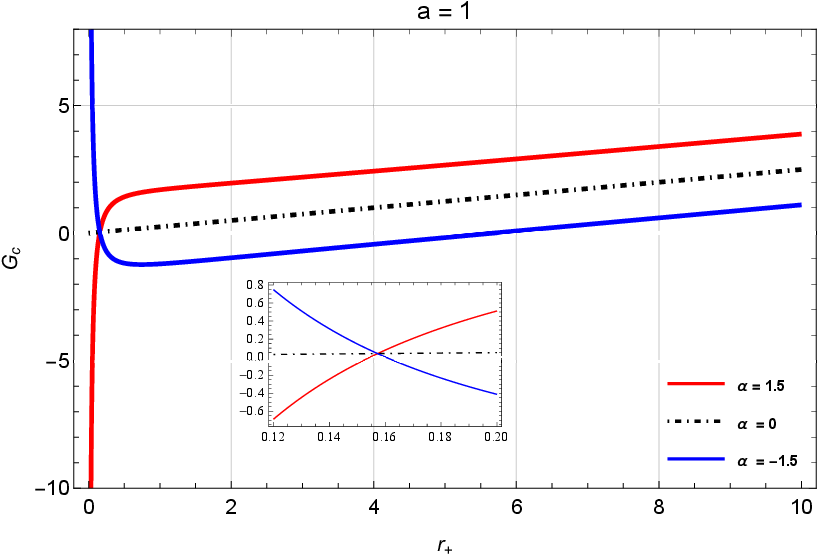}\label{7a}}
\subfigure[]{\includegraphics[width=0.52\linewidth]{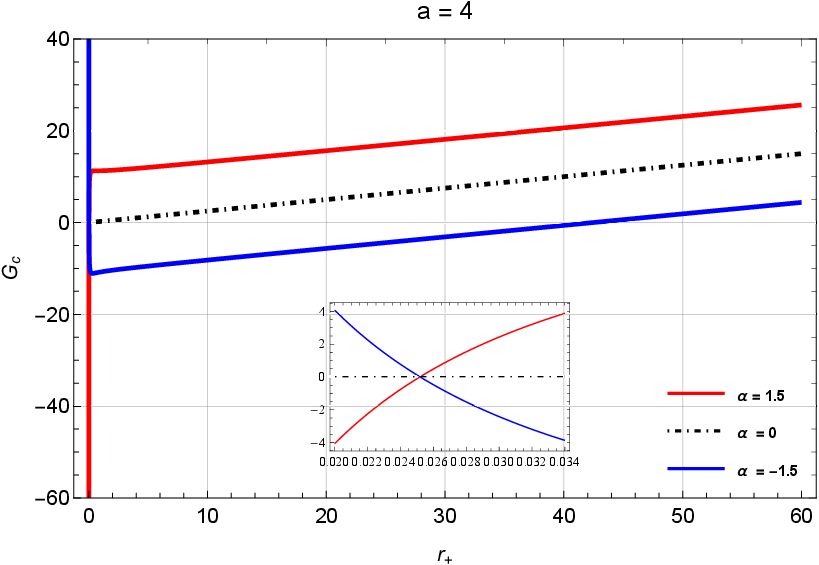}\label{7b}}\\ 
\subfigure[]{\includegraphics[width=0.52\linewidth]{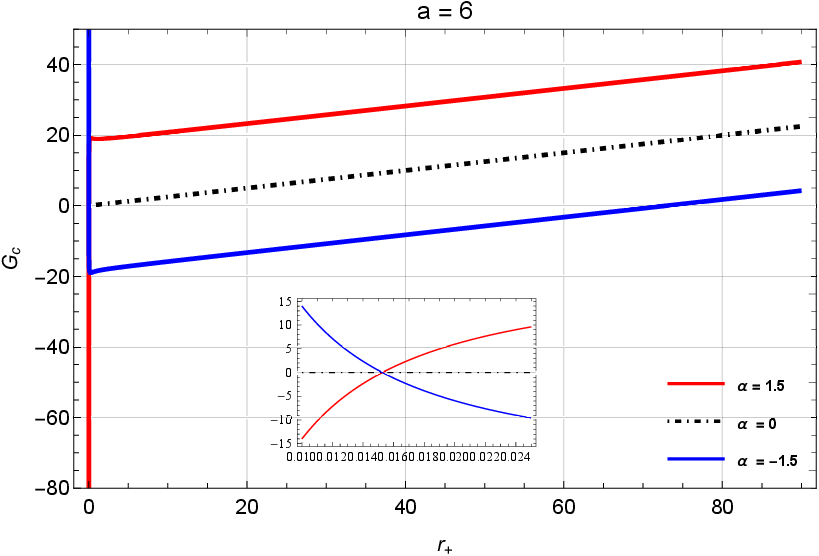}\label{7c}}
\end{array}$
\end{center}
\caption{Corrected Gibbs free energy vs. black hole horizon radius $r_{+}$. Here, $\alpha = 0$ (no correction) is represented by a black dash-dot line, $\alpha = 1.5$ (with correction) is denoted by the red curve, and $\alpha = -1.5$ (with correction) is denoted by the blue curve. Here, $\Lambda=10^{-123}$.}
\label{fig7}
\end{figure}

We will now find the effects of thermal fluctuations on Gibbs free energy. In thermodynamics, Gibbs free energy measures the maximum amount of mechanical work. It is mathematically determined by the given relation
\begin{equation}\label{36}
G_{c}=F_{c}+PV_{c}.
\end{equation}
Inserting the corrected values of Helmholtz free energy and volume we can get
\begin{multline}\label{37}
G_{c}=\frac{1}{4 \pi  r_+} \Bigg[ 8 a \alpha  r_+ \left\lbrace\ln \left(4 a r_+-3 \Lambda  r_+^2+1\right)-\ln \left(r_+\right)\right\rbrace+\alpha  \ln (16 \pi ) \left(3 \Lambda  r_+^2-1\right)\\+\alpha  \left(1-3 \Lambda  r_+^2\right) \ln \left\lbrace\left(4 a r_+-3 \Lambda  r_+^2+1\right){}^2\right\rbrace+r_+^2 \left(24 \alpha  \Lambda -\pi  \Lambda  r_+^2+\pi \right) \Bigg].
\end{multline}
%M\ln|r_{+}|-\frac{\Lambda r^3_{+}}{6}+\frac{\Lambda\alpha r_{+}}{\pi}+\frac{\alpha}{2\pi}\Bigg[-2\Big(\Lambda r_{+}-\frac{M}{r_{+}^2}\Big)\ln(M-\Lambda r_{+}^3+ar_{+}^2)+\Big(\Lambda r_{+}-\frac{M}{r_{+}^2}\Big)\ln 4\pi r_{+}^2\\-8\Lambda r_{+}+4a\ln|r_{+}|-2 a\ln(M-\Lambda r_{+}^3+ar_{+}^2)-\frac{M}{r_{+}^2}\Bigg]
We thus get an expression for corrected Gibbs free energy. The comparison study of corrected and uncorrected Gibbs free energy against the event horizon radius has been shown in fig.\ref{fig7}.

From the plot, we see that Gibbs free energy undergoes a momentous change in its behavior when $r_{+}$ goes to zero. We found one critical point for a small value of $r_{+}$. Before the critical point, one can observe that Gibbs free energy creates the positive (negative) asymptotic value for the negative (positive) correction parameter. The negative value of Gibbs free energy (in the region before the critical point) for correction parameter $\alpha=1.5$ illustrates the sign of stability and indication of the maximum amount of energy that can be turned out from the thermodynamical system of RMSBH. \textcolor{black}{Also beyond the critical point, we observe that Gibbs free energy curves exhibit the same trend as in Helmholtz free energy graph}. Moreover, like Helmholtz free energy, the critical point for $G_{c}$ moves towards lower $r_{+}$ for higher $a$.

\section{ Stability of Rindler modified Schwarzschild black hole}\label{sec5}

In order to study the stability of RMSBH we investigate the nature of its specific heat. From the behavior of specific heat, we can explore whether this black hole undergoes phase transition or not. The positive value of specific heat ensures that the given black hole system is stable against the phase transition, however, the negative value of specific heat concludes the instability of the system. By taking thermal fluctuation into account, we estimate the expression for specific heat which must subside to the uncorrected specific heat (when fluctuation is switched off i.e. $\alpha=0$). From a classical thermodynamics point of view, the specific heat ($C_{c}$) can be calculated with the help of the following standard formula: 
\begin{equation}\label{38}
C_{c}=T_{H}\frac{dS_{c}}{dT_{H}}.
\end{equation}

Substituting the values of Hawking temperature from Eq.(\ref{5}) and the corrected entropy from Eq.(\ref{28}) in Eq.(\ref{38}), we can easily arrive at the expression for leading-order corrected specific heat for our black hole as
\begin{equation}\label{39}
C_{c}=\frac{2 r_+ \left(4 a \alpha +\pi  r_+ \left(-4 a r_++3 \Lambda  r_+^2-1\right)-6 \alpha  \Lambda  r_+\right)}{3 \Lambda  r_+^2+1}.
\end{equation}

\begin{figure}[ht]
\begin{center} 
$\begin{array}{cccc}
\subfigure[]{\includegraphics[width=0.52\linewidth]{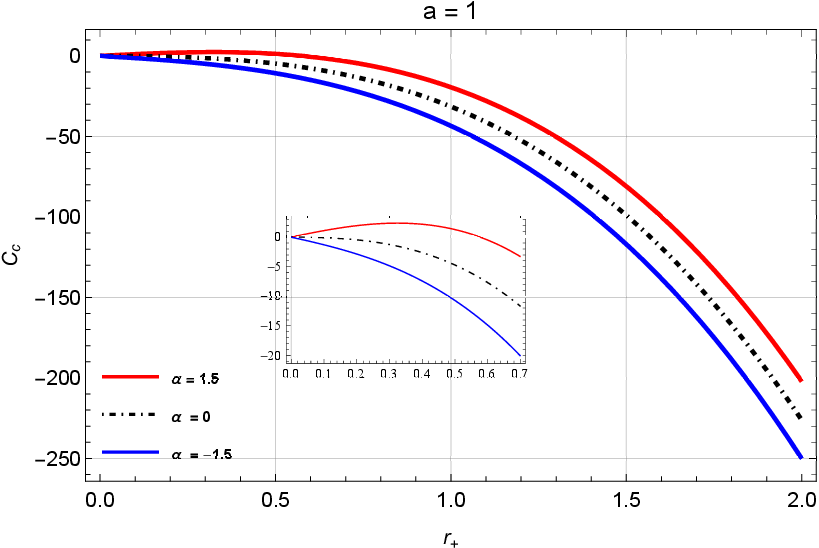}\label{8a}}
\subfigure[]{\includegraphics[width=0.52\linewidth]{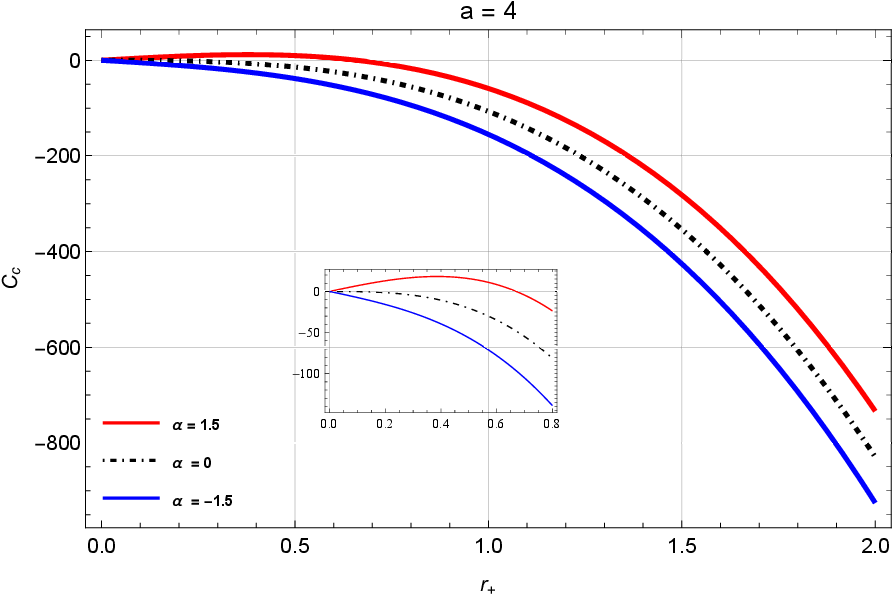}\label{8b}}\\
\subfigure[]{\includegraphics[width=0.52\linewidth]{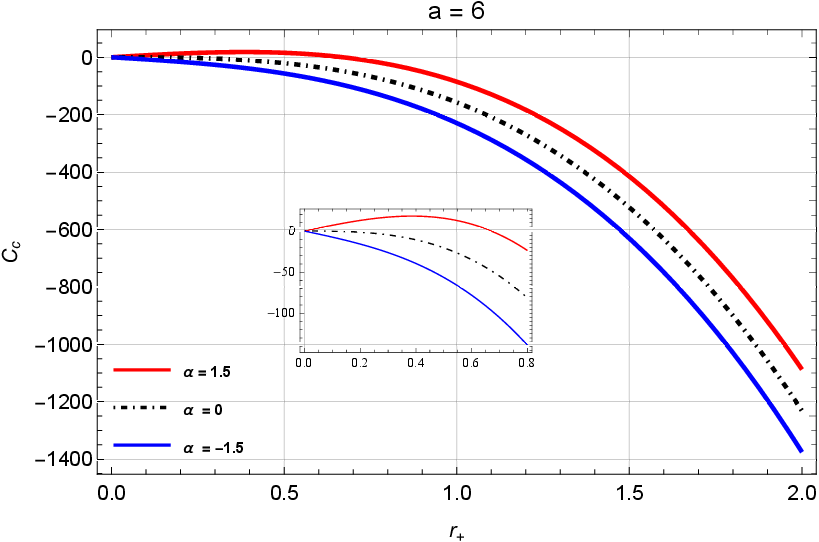}\label{8c}}
\end{array}$
\end{center}
\caption{Corrected specific heat vs. black hole horizon radius $r_{+}$. Here, $\alpha = 0$ (no correction) is represented by a black dash-dot line, $\alpha = 1.5$ (with correction) is denoted by the red curve, and $\alpha = -1.5$ (with correction) is denoted by the blue curve. We set $\Lambda=10^{-123}$.}
\label{fig8}
\end{figure}

We plot (Fig.\ref{fig8}) the corrected specific heat (Eq.(\ref{39})) vs. the event horizon radius graph for fixed values of $\Lambda$ to see the effect of small statistical fluctuations on the stability of our black hole system.

If thermal fluctuations are not taken into consideration (i.e. $\alpha=0$), the specific heat curve starts with zero value for small black holes and then goes on increasing in negative regions for large-sized black holes. This indicates that without any thermal fluctuation, the small-sized black hole is in equilibrium and the large-sized black hole is unstable.  When we consider the thermal fluctuations around the equilibrium, then the specific heat with a positive value of the correction parameter becomes positively valued for the small-sized black hole and becomes negatively valued for the large-sized black hole. It means due to thermal fluctuations for positive correction parameter, the small-sized black hole is stable and the large-sized black hole is unstable. For both the small and large black hole cases, the specific heat curve becomes negatively valued for the negative value of the correction parameter which in turn gives the unstable behavior for our black hole system. 

\textcolor{black}{
Now, finally we discuss the first law of black hole thermodynamics for our black hole system. The first law of thermodynamics is given by \cite{m15}:
\begin{equation}\label{t1}
dM=TdS+VdP
\end{equation}
where, mass of the black hole $M$ is equivalent to its enthalpy energy $H_c$ \cite{rev01}. For this scenario, one can find,
\begin{equation}
    TdS+VdP = \Bigg[\frac{3 \alpha  \Lambda }{\pi }-\frac{2 a \alpha }{\pi  r_+}+2 a r_+-\frac{1}{2} 3 \Lambda  r_+^2+\frac{1}{2} \Bigg] d r_+ = dH_c.
\end{equation}
Hence, in the presence of thermal fluctuations, the first law of black hole thermodynamics is valid.
}

\section{Summary and final remarks}\label{sec7}
Now, we summarize and conclude the final remarks of our work. In this work, motivated by entropy-area law, we have considered the Rindler-modified Schwarzschild black hole and discussed their thermodynamics. Maintaining a close analogy between classical thermodynamics and black hole thermodynamics, we first compute various thermodynamical parameters for this black hole system. For illustration, starting from Hawking temperature and entropy for RMSBH, we derived enthalpy, Helmholtz free energy, pressure, volume, internal energy, and Gibbs free energy of the system in equilibrium. In order to get an answer to
the question of what happens to the stable thermal system when small fluctuations around the equilibrium are taken into account, we computed the leading-order corrections to the entropy of the considered black hole. 

\textcolor{black}{ We found that quantum correction to entropy has importance only when a black hole has a very small horizon radius. A positive value of the correction parameter increases the entropy drastically in comparison to the uncorrected entropy curve for $r_+$ smaller than the critical point. Beyond this point, the corrected entropy is slightly lower than the uncorrected one. The deviation from the uncorrected entropy for large black holes increases for large values of $a$. An increase in $a$ also shifts the critical point towards smaller values of $r_+$.}

Once we derive the Hawking temperature and modified entropy, we then calculated various thermodynamical potentials of RMSBH considering the effect of thermal fluctuations. In this context, the leading-order corrected enthalpy energy of the given black hole system has been estimated. Graphical behavior shows that thermal fluctuations will be significant only when a black hole has a small horizon radius whereas the enthalpy for large black holes will be unaffected. For black holes having small event horizon radius, the enthalpy takes a negative value for the negative correction parameter. However, it takes a positive value corresponding to the positive correction parameter. For higher values of Rindler acceleration ($a$), the enthalpy increases significantly but it does not affect the critical point. Furthermore, to compute the possible amount of energy remaining for doing work, we then derived Helmholtz free energy. To study the dependency of free energy on thermal fluctuations, we present the Helmholtz free energy versus the event horizon radius graph. The figure depicts only one critical point for free energy. For a black hole whose horizon radius is smaller than the critical point, free energy with a negative correction parameter takes the positive asymptotic value while free energy with a positive correction parameter takes the negative asymptotic value. Our analysis shows that quantum corrections affect the thermodynamics of RMSBH as well as other black hole systems significantly at small horizon radii. Here, the role of the larger Rindler acceleration parameter is to move the critical point slowly toward the lower horizon radius. We then analyzed the effect of quantum (thermal) fluctuations on the volume. Moreover, once we derive the Hawking temperature, enthalpy, entropy, pressure, and volume, we calculated the leading-order corrections to internal energy and Gibbs free energy. Nevertheless, to understand the effect of quantum fluctuations on the stability of the black hole, we study the behavior of corrected specific heat. If thermal fluctuations are not taken into consideration (i.e. $\alpha = 0$), the specific heat curve starts with zero value for small black holes and then goes on increasing in negative regions for large-sized black holes. This indicates that without any thermal fluctuation, the small-sized black hole is in equilibrium and the large-sized black hole is unstable. The specific heat with a positive value of the correction parameter becomes positively valued for the small-sized black hole and becomes negatively valued for the large-sized black hole as far as the thermal fluctuations of the black hole system are concerned. It means due to thermal fluctuations for positive correction parameter, the small-sized black hole is stable and the large-sized black hole is unstable. For both the small and large black hole cases, the specific heat curve becomes negatively valued for the negative value of the correction parameter which in turn gives the unstable behavior for our black hole system.

In a pertinent study documented in Ref. \cite{r2}, it was ascertained that the phenomenon of gravitational lensing for the RMSBH consistently yields values lower than those for the corresponding Schwarzschild black hole. Additionally, it was established that the degree of deflection angle exhibited by gravitational lensing diminishes as the parameter $a$ is increased. Similar to these findings, our investigation reveals a notable impact of the parameter $a$ on the specific heat behavior. Specifically, as the value of $a$ is augmented, there is a discernible decrease in the overall stability of the black hole under consideration.

When considering a positive correction parameter, a rather intriguing revelation emerges. It becomes evident that the realm of stable black holes is confined to an exceedingly narrow domain characterized by minute values of $r_+$. This intriguing finding implies that the presence of the Rindler acceleration parameter serves as an indicative marker for the potential existence of diminutive black holes within the cosmic landscape. Such smaller black holes, as concluded in Ref. \cite{r2} can have smaller deflection angle or smaller gravitational lensing.

In the context of future research endeavours, a compelling avenue to explore involves delving into the influence exerted by quantum (thermal) fluctuations upon the critical characteristics of the $P-V$ behavior. This exploration can be undertaken by treating the RMSBH as an analogous system to a Van der Waals fluid. It is noteworthy that within the framework of $f(R)$ gravity, the RMSBH metric demonstrates remarkable efficacy.

Moreover, directing our gaze toward prospective investigations, it becomes enticing to contemplate the extension of these findings to encompass other variants of modified gravity theories. A particularly intriguing avenue involves probing the intricate realm of perturbations within the spacetime of the black hole. Furthermore, a comprehensive scrutiny of the optical traits inherent to black holes, such as the enigmatic black hole shadow, while factoring in the presence of the Rindler acceleration, holds promise for illuminating us with profound insights \cite{Vagnozzi:2022moj}. Such investigations have the potential to offer a more lucid comprehension of the multifaceted repercussions associated with the Rindler acceleration parameter.

\section*{Acknowledgment}
SM, SD, and AP thank A. Dey (Department of Physics, Jadavpur University, India) for suggestions in the early stage of the work. DJG acknowledges the contribution of the COST Action CA21136  -- ``Addressing observational tensions in cosmology with systematics and fundamental physics (CosmoVerse)". We are also grateful to Prof. Rabin Banerjee (Raja Ramanna Fellow, Department of Astrophysics and High Energy Physics, S.N. Bose National Centre for Basic Sciences, India) for his valuable comments on our manuscript. We are thankful to the anonymous referees for their constructive comments.

\section*{Declaration of competing interest}
The authors declare that they have no known competing financial interests or personal relationships that could have appeared to influence the work reported in this manuscript.

\section*{Data availability}
Data sharing is not applicable to this article, as no data sets were analyzed or generated during the current study.

\end{document}